\documentclass[superscriptaddress,showpacs,twocolumn,floatfix,amsmath,footinbib,amssymb]{revtex4-1}
\usepackage{amssymb}
\usepackage{amsmath}
\usepackage{epsfig}
\usepackage{t1enc}
\usepackage{soul}
\usepackage{color}
\usepackage{verbatim}
\usepackage{graphicx}
\usepackage{layout}
\newcommand{\beq}{\begin{equation}}
\newcommand{\eeq}{\end{equation}}
\newcommand{\bea}{\begin{eqnarray}}
\newcommand{\eea}{\end{eqnarray}}

\begin{document} 
\raggedbottom 
\title{Gravitational parameter estimation in a waveguide}
\author{Jason Doukas}
\affiliation{School of Mathematical Sciences, University of Nottingham, Nottingham NG7 2RD, United Kingdom}
\author{Luke Westwood}
\affiliation{School of Mathematical Sciences, University of Nottingham, Nottingham NG7 2RD, United Kingdom}
\author{Daniele Faccio }
\affiliation{School of Engineering and Physical Sciences, Heriot-Watt University, EH14 4AS 
Edinburgh, UK}
\author{Andrea Di Falco }
\affiliation{School of Physics and Astronomy, University of St. Andrews, North Haugh, St. Andrews KY16 9SS, UK}
\author{Ivette Fuentes}
\affiliation{School of Mathematical Sciences, University of Nottingham, Nottingham NG7 2RD, United Kingdom}
\date{\today}
\begin{abstract}
We investigate the intrinsic uncertainty in the accuracy to which a static spacetime can be measured from scattering experiments. In particular, we focus on the Schwarzschild black hole and a spatially kinked metric that has some mathematical resemblance to an expanding universe.  Under selected conditions we find that the scattering problem can be framed in terms of a lossy bosonic channel, which allows us to identify shot-noise scaling as the ultimate scaling-limit to the estimation of the spacetimes. Fock state probes with particle counting measurements attain this ultimate scaling limit and the scaling constants for each spacetime are computed and compared to the practical strategies of coherent state probes with heterodyne and homodyne measurements. A promising avenue to analyze the quantum-limit of the analogue spacetimes in optical waveguides is suggested.
\end{abstract}
\pacs{03.65.Ud, 03.30.+p, 03.67.-a, 04.62.+v}
\maketitle

\section{Introduction}
Heisenberg's uncertainty principle is one of the defining features of measurement that distinguishes quantum and classical mechanics. It imposes a fundamental limit to the precision with which non-relativistic complementary variables, like position and momentum, can be known simultaneously. As argued by Unruh \cite{Unruh1986} the limits imposed by quantum measurement in gravitational contexts can lead to surprising insights that may even shed light on the problem of the quantization of gravity.  A better understanding of these quantum imposed limitations on the measurability of space-time is therefore of great interest. 

Understanding how the measurement precision of a physical quantity scales with the available resources is currently a very active area of research going by the name of quantum metrology (see the reviews \cite{Giovannetti2011, Giovannetti2004}). Quantum metrology has been used for ultra precise gravitational wave searches \cite{Caves1981,LIGOcollab}, frequency calibration in atomic spectroscopy \cite{Spectroscopy}, sub-classical quantum lithography \cite{DAngelo2001, Theoryquantlith} and entanglement-assisted magnetometry \cite{Wasilewsk} and electrometry \cite{Sedlacek} to name a few.  

Measuring physical quantities that play a role in relativity such as gravitational field strengths, proper accelerations and space-time parameters is of great interest not only to science but also to technology.  Recently, techniques that apply quantum metrology to quantum field theory in curved and flat space-time have been developed \cite{Aspachs, Downes, RQM, RQM2}. The application of these techniques can in principle produce technologies that outperform non-relativistic quantum estimation of gravitational parameters. Indeed, it was shown that relativistic effects, such as particle creation, can be exploited to improve the measurement of accelerations \cite{RQM} and the detection of gravitational waves \cite{Sabin}. Earlier work showed that the mode entanglement generated by the expansion of the universe encodes the expansion rate of the universe~\cite{Ball}. Also is was shown that phase estimation techniques could be employed to measure the Unruh effect at accelerations that are within experimental reach~\cite{Aspachs} (see also \cite{Doukas2013} for the application of channel discrimination to such experiments).  

A typical problem in quantum metrology is that of quantum parameter estimation whereby one attempts to find the best estimation of an unobservable continuous variable, $\theta$, that parameterizes a state, $\rho(\theta)$. An illustrative and pertinent example is the two-mode beam splitter. When a fixed state of light is shone onto a beam splitter, the beam splitter reflectivity is encoded into the output state. Since there is no reflectivity observable, no measurement exists that determines the reflectivity precisely. Rather the reflectivity must be inferred by performing other measurements on the output state. By performing a general positive operator-valued measurement $\{\hat{O}_x\}$, statistical techniques \cite{Cramer, Helstrom, Holevo} can be applied to the obtained probability distribution, $p(x|\theta)=\text{Tr}[\hat{O}_x\rho(\theta)]$, to estimate the reflectivity of the beam splitter. For an unbiased estimator, and $N$ repetitions of the experiment, the variance of the parameter is bounded by the Cramer-Rao inequality, $(\Delta\theta)^2\geq1/NF(\theta)$, where 
\bea\label{FI}
F(\theta)\equiv\int p(x|\theta)\left(\frac{d \log {p(x|\theta)}}{d\theta}\right)^2dx,
\eea
is the Fisher information. The Fisher information is bounded above by the quantum Fisher information which provides a measure of the ultimate precision attainable (i.e., best measurement strategy) for a given probe state. Since the quantum Fisher information is achievable asymptotically \cite{Braustein1996}, the quantum Cramer-Rao bound provides a parameter based uncertainty relation for the unobservable quantity.

Recently, several examples of the optimal quantum Cramer-Rao lower bound for measuring various single parameter spacetime metrics have been given \cite{Downes} (see also \cite{Aspachs,RQM, RQM2, Sabin, Unruh1986, Wang}). These relations were found by applying the abstract formulation of quantum field theory in curved spacetime known as the locally covariant approach \cite{locallcov}.  In this paper, we instead find quantum-limitations on the spacetime measurement by investigating scattering experiments in the usual formalism of quantum field theory in curved spacetime \cite{Birrell1982}. 

Scattering experiments have proved extremely fruitful in probing the details of subatomic phenomena \cite{Schroeder, Sakurai} and are also commonly used for measuring properties of larger bodies, for example in radar. The theory of scalar wave scattering from black holes has a long history \cite{scalarBHscatter} (see \cite{bhscatterbook} for a comprehensive review).  Until now these works have invariably focussed on measurements of the differential cross-section of scattered plane-waves. This setup is principally classical because the problem can always be phrased in terms of classical waves and intensity measurements.  We consider here whether the strategy of scattering quantum probes and performing quantum measurements (i.e., homodyne, heterodyne, particle counting etc) can attain greater precision on the information of gravitational objects, like black holes. This is motivated by the fact that at fixed energy quantum metrological strategies generally give heightened sensitivity over those which are purely classical. While the quantum treatment of fields on gravitational backgrounds is not new (indeed Hawking radiation \cite{Hawking1975} itself was derived by considering the quantum nature of the field that is scattered through a collapsing body) to our knowledge the quantum scattering problem has not been analyzed in the context of precision measurements of the spacetime.   

We find the precision sensitivity limits for the space-time parameters as a function of the given energy resources and find that because the channel is intrinsically noisy the variance is limited by the classical shot-noise scaling. While the Heisenberg limit (a quadratic improvement over the shot-noise scaling) cannot be achieved in this setup we present the optimal constant scaling factor and compare it against other strategies.
  
At optical wavelengths optimal estimation occurs for black holes with masses that are too small to be produced by any known physical processes. Rather we find that for black holes with masses of the order of magnitude of the sun radio-waves are required. This rules out a standard quantum optics implementation for physical black holes. Nevertheless, using a correspondence between the propagation of electromagnetic fields in dielectric waveguides, we propose a quantum optics experiment capable of verifying the general concepts we have outlined for micron-sized \textit{analogue} black holes.  
 
\section{Black hole scattering}\label{sec:bhscattering}

In our first example we consider quantum scattering experiments in the exterior region of the Schwarzschild black hole. The Schwarzschild metric is:
\bea
ds^2=-f_M(r)dt^2+f_M(r)^{-1}dr^2+r^2 d\Omega^2,
\eea
where $f_M(r)=1-2M/r$ and $d\Omega^2=d\theta^2+\sin^2{\theta}d\phi^2$. Scalar field perturbations satisfy the Klein-Gordon equation on the unperturbed Schwarzschild background:
\beq
\frac{1}{\sqrt{-g}}\partial_{\mu}(g^{\mu\nu}\sqrt{-g}\partial_{\nu} \phi)=0,
\eeq
with the solution $\phi=\sum_{l,m} \int d\omega r^{-1}R(r)Y_{l,m}(\theta,\phi)e^{-i\omega t}$, where $Y_{l,m}$ are spherical harmonics and the radial function, $R(r)$, satisfies the equation:
\bea\label{masterequation}
\frac{d^2 R}{dx^{2}}+(\omega^2-V(r))R=0,
\eea
where  $x=r+2M\log(r/2M-1)$, and:
\bea\label{potential}
V(r)=f(r)\left(\frac{l(l+1)}{r^2}+\frac{2M}{r^3}\right).
\eea 
The first term in the potential comes from the centrifugal barrier, while the second is due to the curvature of the spacetime \footnote{As we are only interested in the qualitative features we use a scalar field to illustrate the approach. However, in reality only the electromagnetic field will be available to perform the scattering. Since equation (\ref{masterequation}) also holds for the electromagnetic field \cite{Wheeler1957, Ruffini} for $l\geq1$ with the same potential but without the curvature term,  our technique can be trivially repeated with the electromagnetic field instead of the scalar field. }. The potential reaches its maximum at some position $x_{\text{max}}$ outside the event horizon and vanishes both at spatial infinity, $x\rightarrow \infty$, and just outside of the horizon $x\rightarrow-\infty$.

We consider an observer at spatial infinity attempting to measure the black hole mass by scattering waves off the potential. For simplicity, we assume that the probe wave is quasimonochromatic having a very narrow frequency spread relative to its midfrequency $\Delta \omega/\omega\ll 1$. For convenience we also suppose that the ingoing waves have definite values of $l$ and $m$ \footnote{It is worth mentioning that $s$-wave scattering is often a good approximation for low energy plane-wave scattering. One could then realize our setup in the case ($l=0$, $m=0$), using low energy plane-waves.}. By the conservation of energy and angular momentum the scattered waves will maintain these properties. We further assume that the black hole mass is large enough that it is not significantly increased by the infalling particles from the probe.  

\begin{figure}
\centering
\includegraphics{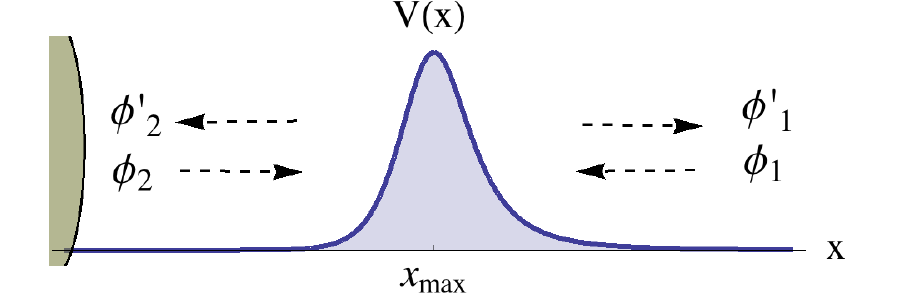}
\caption{\label{fig:potential} (Color online) Plot of the potential (blue filled curve) in the exterior black hole region; the even horizon is at $x\rightarrow -\infty$. The modes are labelled with subscripts $1$ and $2$, where $1$ indicates the mode is either entering (unprimed) or exiting (primed) the scattering center from the horizon ($x\ll x_{\text{max}}$) while the subscript $2$ indicates that the mode is either entering (unprimed) or exiting (primed) the scattering center from outside the potential ($x\gg x_{\text{max}}$). }
\end{figure}

In general, exact solutions to (\ref{masterequation}) do not exist, however near the horizon and at spatial infinity the radial solutions take the asymptotic form $R\sim e^{\pm i\omega x}$, see Fig.~\ref{fig:potential}. The ``entering'' modes behave like $\phi_1\sim e^{i\omega (x-t)}$ as $x\rightarrow -\infty$ and $\phi_2\sim e^{-i\omega( x+t)}$ as $x\rightarrow \infty$.  Similarly, the ``exiting'' modes, denoted by primes, behave like $\phi_1'\sim e^{-i\omega (x+t)}$ as $x\rightarrow -\infty$ and $\phi_2'\sim e^{i\omega (x-t)}$ as $x\rightarrow \infty$. One easily checks (using the Wronskian) that $\phi_1$ \& $\phi_1'$ and $\phi_2$ \& $\phi_2'$ are respective pairs of linearly independent solutions. By evolving the $\phi_1$ and $\phi_1'$  modes through to the other side of the scattering potential, one is able to determine the transfer matrix, $M$,  which relates the $1$-modes to the $2$-modes:
\bea
\left(\begin{array}{c} \phi_1\\
\phi_1'\end{array}\right)=M\left(\begin{array}{c} \phi'_2\\
\phi_2\end{array}\right).
\eea
Using $R'_1(x\rightarrow-\infty)=R^*_1(x\rightarrow-\infty)$ and $R'_2(r^*\rightarrow\infty)=R^*_2(r^*\rightarrow\infty)$, we find the relations between the transfer matrix elements: $M_{21}=M_{12}^*$ and $M_{22}=M_{11}^*$. Also, since the 3-current of a stationary solution in a time-independent potential is divergenceless, one can evaluate the current flux-integral through the surfaces of two concentric spherical shells at $x\rightarrow\pm\infty$ to find $\text{det}(M)=1$.  
The unitary scattering matrix, $S$, which takes unprimed ``entering'' modes into primed ``exiting'' modes, is then obtained by simple rearrangement:
\bea\label{Smatrix}
\left(\begin{array}{c} \phi_1'\\
\phi_2'\end{array}\right)=S\left(\begin{array}{c} \phi_1\\
\phi_2\end{array}\right); \quad S=\frac{1}{M_{11}^*}\left(\begin{array}{cc} M_{12} & 1\\ 1 & -M_{12}^*\end{array}\right).
\eea
Everything until this point has been a purely classical wave calculation. In order to introduce quantum states the field must be quantized. Quantization of fields on static spacetimes is now well-understood and generally believed to be correct \cite{Fulling1972, Birrell1982}. To define the annihilation operators for the ``entering'' and ``exiting'' modes one uses the Klein-Gordon inner product $(f_1,f_2)=i\int f_1^*\stackrel{\leftrightarrow}{\partial^{\mu}}f_2 d\Sigma_{\mu}$, where $\Sigma$ is a constant-$t$ hypersurface \footnote{We use the convention of anti-linearity in the first argument.}. Then the positive frequency modes ($\omega>0$) result in the annihilation operators for $i=1,2$:
\bea
\hat{a}'_i&\equiv&\big(\phi'_i,\hat{\Phi}\big);\quad \hat{a}_i\equiv\big(\phi_i,\hat{\Phi}\big),
\eea
where $\hat{\Phi}$ is the field operator. From the orthonormality of the $\phi_1$ and $\phi_2$ field modes, the annihilation operators can be seen to satisfy the usual commutation relations: $[\hat{a}_1,\hat{a}^{\dagger}_1]=[\hat{a}_2,\hat{a}^{\dagger}_2]=1$ (and similarly for primed modes). Then equation (\ref{Smatrix}) reveals that the scattering is a two-mode passive unitary channel: $\hat{a}'_i=S_{ij}^*\hat{a}_j$. For scattering from outside the potential $S_{11}^*$ and $S_{12}^*$ are the reflection and transmission amplitudes respectively. Reparameterizing $S_{11}^*=e^{i\theta_R}\cos{\phi}$ and $S_{12}^*=e^{i\theta_T}\sin{\phi}$ we obtain:
\bea\label{Sstar}
S^*=\left(\begin{array}{cc} e^{i\theta_R} & 0\\
0 & -e^{i(\theta_T+\theta)}\end{array}\right)\left(\begin{array}{cc} \cos{\phi}& e^{i\theta}\sin{\phi} \\
-e^{-i\theta}\sin{\phi}  &  \cos{\phi}\end{array}\right),
\eea
where $\theta\equiv \theta_T-\theta_R$. The black hole scatterer is carrying out the same operation as a two-mode beam splitter followed by two single mode phase-shift operations. 

We choose here to investigate the strategy of a single observer far away from the black hole \footnote{While one could also consider the situation of a second observer performing local measurements near the horizon we find this setup less realistic due to the tidal forces that such an observer would have to endure.}. As mode $2$ is not measured, the phase-shift can be absorbed i.e., $\hat{a}'_2\rightarrow -e^{-i(\theta_T+\theta)} \hat{a}'_2$. On the other hand, the $e^{i\theta_R} $ phase-shift is measurable if the observer at infinity keeps a very precise local clock \footnote{Phase-shifts in scattering experiments are common and occur for example in the outgoing partial waves of plane-wave scattering from spherically symmetric potentials \cite{Sakurai}. Typically such phase-shifts are estimated by performing a $\chi^2$-fit to the measured differential cross-section, see for example \cite{Andrick}.  Our case is analogous to $s$-wave scattering in which only a single phase-shift occurs. The crucial difference in black holes is that part of the state is transmitted by the potential and eventually absorbed by the black hole. In plane-wave scattering this would manifest as a complex phase shift in the partial waves of the differential cross-section and is categorized as ``inelastic'' scattering.} and the signal does not drift over the duration of the experiment.  This latter requirement is somewhat unrealistic when one considers astrophysical timescales. Nevertheless, it is still possible to phase-lock the signal to a clock by sending the phase reference (local oscillator) through the potential along with the signal. However, then both the signal and local oscillator receive the same phase-shift which then becomes unobservable (i.e., their relative phase is fixed).  This situation can be expressed by redefining the operator $\hat{a}'_1\rightarrow -e^{-i\theta_R} \hat{a}'_1$, then the phase-locked scattering corresponds to only beam-splitting. Thus with these redefinitions of the mode operators the effective transformation on the entering modes becomes simply the right matrix on the r.h.s of equation (\ref{Sstar}).  

We assume that the observer, who is far from the black hole, can only measure the reflected part of the wave. The partial loss of the initial state due to the transmission into the black hole makes the scattering as observed from infinity non-unitary. To see this we note that since the mode $\hat{a}_2'$ gets lost in the black hole, the map acting on the input state $\rho=\rho_1\otimes|0\rangle_2\langle 0|_2$, is given by operating with the unitary representation of the beam splitting operation $D(\chi)$ and then tracing over mode 2:
\bea
\mathcal{E}(\rho)\equiv\text{Tr}_2\left[D(\chi) \rho_1\otimes |0\rangle_2 \langle 0|_2 D^{\dagger}(\chi)\right],
\eea
where $D(\chi)=\exp{(\chi \hat{a}_1^{\dagger}\hat{a}_2-\chi^* \hat{a}_1\hat{a}_2^{\dagger})}$ with $\chi=\phi e^{i \theta}$. Explicitly, the mode operators transform as:
\bea
\hat{a}_1'&=&D^{\dagger}(\chi)\hat{a}_1D(\chi)=\hat{a}_1 \cos{\phi}+\hat{a}_2 e^{i\theta} \sin{\phi},\label{D1}\\
\hat{a}_2'&=&D^{\dagger}(\chi)\hat{a}_2D(\chi)=-\hat{a}_1e^{-i\theta} \sin{\phi}+\hat{a}_2  \cos{\phi}.\label{D2}
\eea
Note that the state of the ingoing mode, $\phi_2$, has been chosen to be the vacuum. While a collapsed black hole will naturally emit radiation in a thermal state at the Hawking temperature, $T=1/8\pi M $ \cite{Hawking1975, Unruh1976}, this radiation is negligible when the frequency of the probe is much larger than the temperature, $\omega \gg 1/8\pi M$ \footnote{Investigating this problem in the regime in which the Hawking radiation is large is the subject of a future work.}.

Using equations (\ref{D1})-(\ref{D2}) and $\langle 0|_2D(\chi)|0\rangle_2=(\cos{\phi})^{\hat{a}_1^{\dagger}\hat{a}_1}$ which follows from the angular momentum operator ordering theorem \cite{Barnett}, one obtains:
\bea\label{lossymap}
\mathcal{E}(\rho)=\sum_{n=0}^{\infty}\frac{\sin^{2n}\phi}{n!} (\cos{\phi})^{\hat{a}_1^{\dagger}\hat{a}_1} \hat{a}_1^n \rho_1 (\hat{a}_1^{\dagger})^n (\cos{\phi})^{\hat{a}_1^{\dagger}\hat{a}_1}.
\eea
Taking the derivative with respect $\phi$ results in the Lindblad equation: $d\mathcal{E}(\rho)/d\phi=\tan{\phi} \mathcal{L}(\hat{a}_1) \mathcal{E}(\rho)$, where $ \mathcal{L}(\hat{a}_1) \rho\equiv 2\hat{a}_1\rho \hat{a}_1^{\dagger}-\hat{a}_1^{\dagger}\hat{a}_1\rho-\rho \hat{a}_1^{\dagger}\hat{a}_1$, which is the standard master equation for a lossy bosonic channel.

The optimal estimation of the loss parameter $\phi$ for this channel has been thoroughly investigated in the literature \cite{Monras2007, Adesso2009, Sarovar2006}. In particular, in \cite{Adesso2009} it was shown that the strategy which realizes the ultimate quantum limit is the $n$ particle Fock state with particle counting measurements (Fock states were also shown to be optimal probe states in the estimation of the Unruh temperature by observers moving with uniform acceleration \cite{Aspachs}). This strategy has a Fisher information of $4n$ and satisfies the quantum Cramer-Rao bound, $\Delta \phi \geq \frac{1}{\sqrt{4 n N}}$, where $N$ is the number of repetitions of the experiment. Since the reflection parameter is only a function of the mass, using the reparameterization property of the Fisher information, $F(M)= F(\phi(M))\left(\frac{d\phi}{dM}\right)^2$,  we obtain the limit on the uncertainty in the mass of the black hole:
\bea\label{MCRB}
\Delta M \geq \frac{1}{\big|\frac{d\phi}{dM}\big|}\frac{1}{\sqrt{4 n N}}.
\eea
The r.h.s scales like the inverse square root of the energy-resource (i.e., particle number) which is often referred to as the \textit{shot-noise limit}. Although the use of quantum-entanglement in unitary channels is known to lead to a quadratic improvement over the shot-noise limit called the \textit{Heisenberg limit}, it has been shown in \cite{DDobrzanski2012} that  shot noise scaling-- up to some proportionality constant-- is optimal for lossy channels. Since our channel is non-unitary we expect shot-noise scaling and according to the arguments in \cite{Adesso2009}, the proportionality constant of (\ref{MCRB}) is optimal.

To investigate how the information about the mass of the black hole is encoded in this loss parameter, one needs explicit functions of the reflection and transmission amplitudes in terms of the black hole mass. While such exact solutions are not expressible in simple closed form, approximate expressions can be found by making the P\"oschl-Teller approximation \cite{Ferrari}. Under this approximation one finds:
\bea
S_{11}^*&=&\frac{\Gamma(-i\omega/\alpha)\Gamma(1+\beta+i\omega/\alpha)\Gamma(-\beta+i\omega/\alpha)}{\Gamma(i\omega/\alpha)\Gamma(1+\beta)\Gamma(-\beta)},\\
S_{12}^*&=&\frac{\Gamma(1+\beta+i\omega/\alpha)\Gamma(-\beta+i\omega/\alpha)}{\Gamma(1+i\omega/\alpha)\Gamma(i\omega/\alpha)},
\eea 
where $\alpha\equiv\sqrt{- \frac{d^2V(x_{\text{max}})}{dx^2}/2V(x_{\text{max}})}$ is the curvature of the potential at its maximum and $\beta\equiv -1/2 +\sqrt{1/4- V(x_{\text{max}})/\alpha^2}$.
 
\begin{figure}
\centering
\includegraphics[scale=.75]{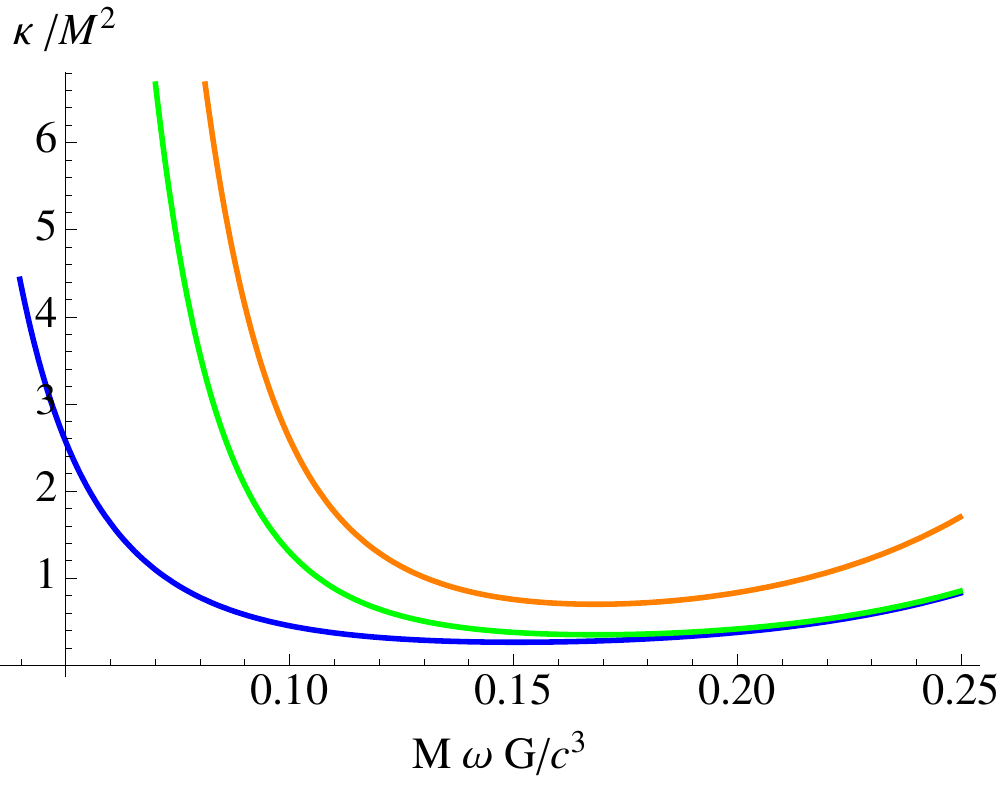}
\caption{\label{fig:cnststrategies} (Color online) Comparison of the constant scaling factor $\kappa$ for different black hole mass estimation strategies. $\kappa$ is defined by the inequality $(\Delta M)^2\geq\frac{\kappa}{n N}$ where $n$ is the mean particle number of the initial state, and $N$ is the number of repetitions of the experiment.  (top orange) Coherent state with heterodyne measurements, (middle green) coherent state with homodyne measurements and (bottom blue) Fock state with particle counting measurements. For all values of the mass the Fock state strategy gives the lowest scaling factor and hence the best estimation of the mass. }
\end{figure}

The results of the optimal experiment are shown in FIG.~\ref{fig:cnststrategies} compared to those of the suboptimal (quantum strategies). The suboptimal examples,  have been chosen because they are very robust experiments that are routinely  performed in optics laboratories (the Fisher information for these strategies is calculated in Appendix \ref{practicalFisher} ). On the other hand precise control over Fock states is still limited to relatively low particle numbers \cite{fock}. 

\begin{figure}
\centering
\includegraphics[scale=.82]{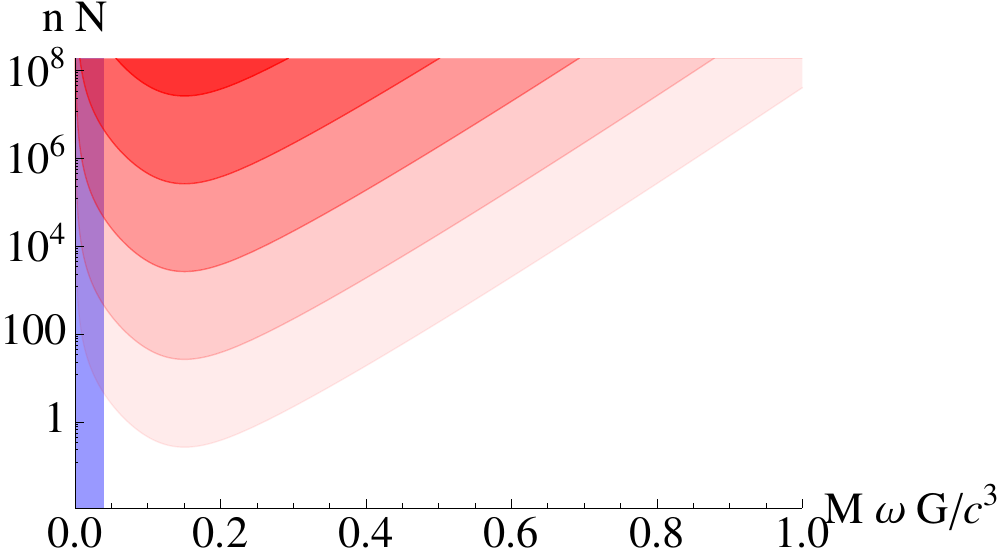}
\llap{\raisebox{1cm}{
\includegraphics[scale=.45]{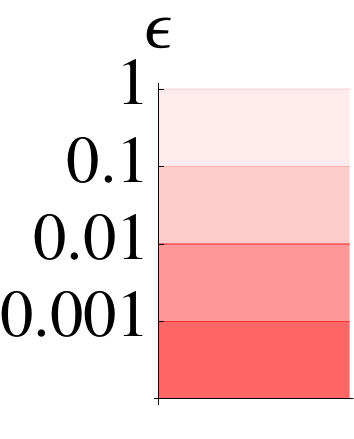}
}}
\caption{\label{fig:refsens} (Color online) Relative sensitivity contour lines as a function of the black hole mass, $n$ is the particle number of the initial Fock state, and $N$ is the number of repetitions of the experiment. Blue shaded region is where Hawking radiation will be significant and where our approximations are no longer valid. }
\end{figure}

In experiments, it is generally harder to achieve a given absolute error when the quantity of interest is of a much larger magnitude than the error itself. A better measure of the quality of a measurement strategy is the relative uncertainty, $\frac{\Delta M}{M}$, which gives the uncertainty or error as a ratio of the quantity itself. We define the relative sensitivity, $\epsilon$, as the minimum relative uncertainty. From equation (\ref{MCRB}) we have:
\bea
\epsilon=\left(M\Big|\frac{d\phi}{dM}\Big|\sqrt{4 n N}\right)^{-1}.
\eea
As shown in Fig.~\ref{fig:refsens}, for a given black hole mass there is an optimal frequency at which the lowest number of resources (particles) are required to obtain the best relative sensitivity of the mass estimation. Therefore, given the order of magnitude of the black hole mass, there is a preferred frequency scale which gives the best estimation of the mass for the lowest number of resources. We find numerically that the optimal frequency is given by:
\bea\label{optimalomega}
\omega_{\text{opt}}=0.15\times \frac{c^3}{MG}.
\eea
This optimal frequency can then be used to determine the best experimental configuration to do black hole mass estimation in an analogue experiment in a waveguide described in more detail in section \ref{sec:waveguide}. 

It is worth checking the consistency of these results in light of the fact that we are neglecting the effects of Hawking radiation. The average number of particles in the frequency band $\omega_{\text{opt}}\pm d\Omega/2$ measured over the pulse duration $d\Omega^{-1}$ is given by the Bose-Einstein factor \cite{Hawking1975}:
\bea
\langle n_{\omega_\text{opt}}\rangle=\frac{1}{e^{8\pi MG\omega_\text{opt}/c^3}-1}.
\eea
Therefore, in such a wave-packet at the optimal frequency (\ref{optimalomega}) we expect only $0.024$ particles from the Hawking effect. This justifies the assumption that the mode is approximately in the vacuum state. 

\section{The kinked spacetime}
We now repeat the previous arguments with an analogue expanding spacetime  which we will refer to as a \textit{kinked spacetime}. We consider a one dimensional static metric,
\bea\label{kinkedmetric}
ds^2=g_{t}(z)dt^2-dz^2,
\eea
 that is asymptotically constant but kinked in the middle according to the profile (see (3.86) of \cite{Birrell1982}):
\bea\label{gtprofile}
-g_t(z(x))=A+B\tanh{\rho x},
\eea
where $x$ is defined in analogy to the tortoise coordinate, $\frac{dz}{dx}=\sqrt{g_t(z)}$, and $A$, $B$ and $\delta$ are constants satisfying $A=-1+\frac{\delta}{2}$ and $B=\frac{\delta}{2}$, $\delta>0$. 
The relationship of (\ref{kinkedmetric}) to the expanding universe metric:
\bea
ds^2=dt^2-g_z(t)dz^2,
\eea
is given \footnote{The analogy should be interpreted here as a symmetry of the equations rather than an emulation of the physics. } by interchange of the spacial and temporal coordinates, $t\leftrightarrow x$, and interchange of the proper-time and proper-distance, $ds^2\leftrightarrow -ds^2$.

A massive scalar field propagating in the geometry (\ref{kinkedmetric}), has the general solution $\phi=Z(z)e^{-i\omega t}$ where $Z$ satisfies the equation:
\bea\label{kinkschrodinger}
\frac{d^2Z}{dx^2}+ (\omega^2-m^2g_t(z))Z=0,
\eea    
which is again of the Schr\"{o}dinger form with a spatially dependent potential this time given by $m^2g_t(z)$. If the potential is parameterized by a single parameter then an observer on one side can estimate the parameter by performing scattering experiments like that described for black holes in the previous section.  The asymptotic form of the solutions to (\ref{kinkschrodinger}) satisfy $Z(x\rightarrow \infty)\sim e^{\pm ik_{2}x}$ and $Z(x\rightarrow -\infty)\sim e^{\pm ik_{1}x}$ where:
\bea
k_{1}&=&\sqrt{\omega^2+(A-B)m^2};\\
 k_{2}&=&\sqrt{\omega^2+(A+B)m^2}.
\eea

For the choice of profile (\ref{gtprofile}) the solutions are \cite{Birrell1982}:
\begin{align}
Z_1'(x)&\equiv \frac{1}{\sqrt{k_1}} \text{exp}{\left(-i k_+ x-i \frac{k_-}{\rho} \ln{(2\cosh{\rho x})}\right)}\times\nonumber\\
_2F_1&\left(1+\frac{ik_-}{\rho},\frac{ik_-}{\rho},1- \frac{ik_{1}}{\rho},\frac{1}{2}(1+\tanh{\rho x})\right),
\end{align}
where $k_{\pm}\equiv \frac{1}{2} (k_{2}\pm k_{1})$. A second solution, is found by: $Z_1(x)=Z_1'(x)^*$.
The other two solutions are given by:
\begin{align}
Z_2(x)&\equiv \frac{1}{\sqrt{k_{2}}} \text{exp}\left({-i k_+ x-i \frac{k_-}{\rho} \ln{(2\cosh{\rho x})}}\right)\times\nonumber\\
_2F_1&\left(1+\frac{ik_-}{\rho},\frac{ik_-}{\rho},1+ \frac{ik_{2}}{\rho},\frac{1}{2}(1-\tanh{\rho x})\right),
\end{align}
and $Z_2'(x)=Z_2(x)^*$. Using (15.3.6) and (15.3.3) of \cite{Abramowitz} the $1$ modes can be related to the $2$ modes via:
\bea\label{z1p}
Z_1'&=&\alpha Z_2 +\beta Z_2',\\\label{z1}
Z_1&=&\alpha^* Z_2' +\beta^* Z_2,
\eea
where 
\bea
\alpha &=& \sqrt{\frac{k_{2}}{k_{1}}} \frac{\Gamma(1-\frac{ik_1}{\rho})\Gamma(-\frac{ik_2}{\rho})}{\Gamma(-\frac{ik_\text{+}}{\rho})\Gamma(1-\frac{ik_\text{+}}{\rho})},\\
\beta &=& \sqrt{\frac{k_{2}}{k_{1}}} \frac{\Gamma(1-\frac{ik_1}{\rho})\Gamma(\frac{ik_2}{\rho})}{\Gamma(\frac{ik_\text{-}}{\rho})\Gamma(1+\frac{ik_\text{-}}{\rho})}.
\eea
By rearrangement of (\ref{z1p})-(\ref{z1}) we we obtain the scattering matrix: 
\bea
\left(\begin{array}{c} \phi_{1}'\\
\phi_{2}'\end{array}\right)=S\left(\begin{array}{c} \phi_{1}\\ \phi_{2}
\end{array}\right);\quad S=\left(\begin{array}{cc}  \frac{\beta}{\alpha^*}& \frac{1}{\alpha^*} \\
 \frac{1}{\alpha^*}&-\frac{\beta^*}{\alpha^*} \end{array}\right). 
\eea
One easily verifies the unitarity from the relation $|\alpha|^2-|\beta|^2=1$. So the scattering matrix is unitary and reduces again into a combination of beam splitting and phase shift operations. Had one taken the expanding space-time analogy too literally one may have been anticipating a squeezing channel (particle creation) instead. Of course this is impossible because the spacetime is static and so the energy is necessarily conserved. However, it is interesting to note that the non-passive squeezing which produces particles in the expanding spacetime has been replaced in the kinked spacetime analogy with a two-mode beam splitting operation. In this sense, particle creation and beam splitting are related under this symmetry. Previous work has shown that entanglement is generated in expanding spacetimes that create particles through a squeezing channel. In such situations, the expansion spacetime parameters can be estimated though the entanglement \cite{Ball}. Here we find the optimal strategy to estimate the parameters of kinked spacetimes where states undergo mode-mixing through a beam splitting channel. 

\begin{figure}
\centering
\includegraphics[width=.5\textwidth]{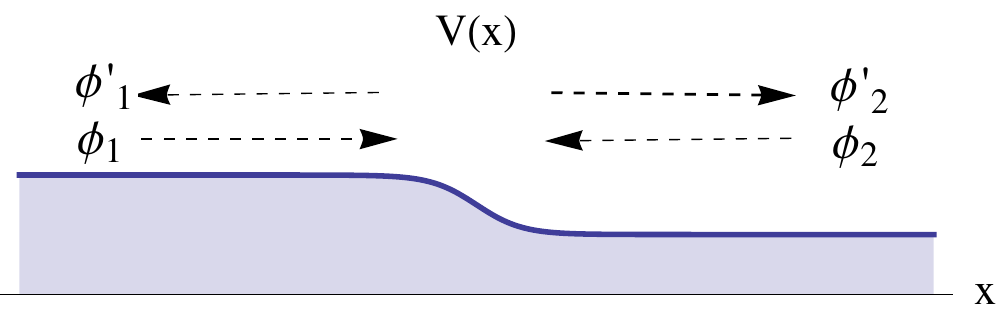}
\caption{\label{fig:potkink} (Color online) Plot of the potential (blue filled curve) for the kinked spacetime. As in the black hole case, the purely ingoing and outgoing modes in the left region where the field is prepared  and measured are labelled with the subscript $1$ while those in the unmeasured region are labelled with the subscript $2$.  }
\end{figure}

\begin{figure}
\centering
\includegraphics[scale=.75]{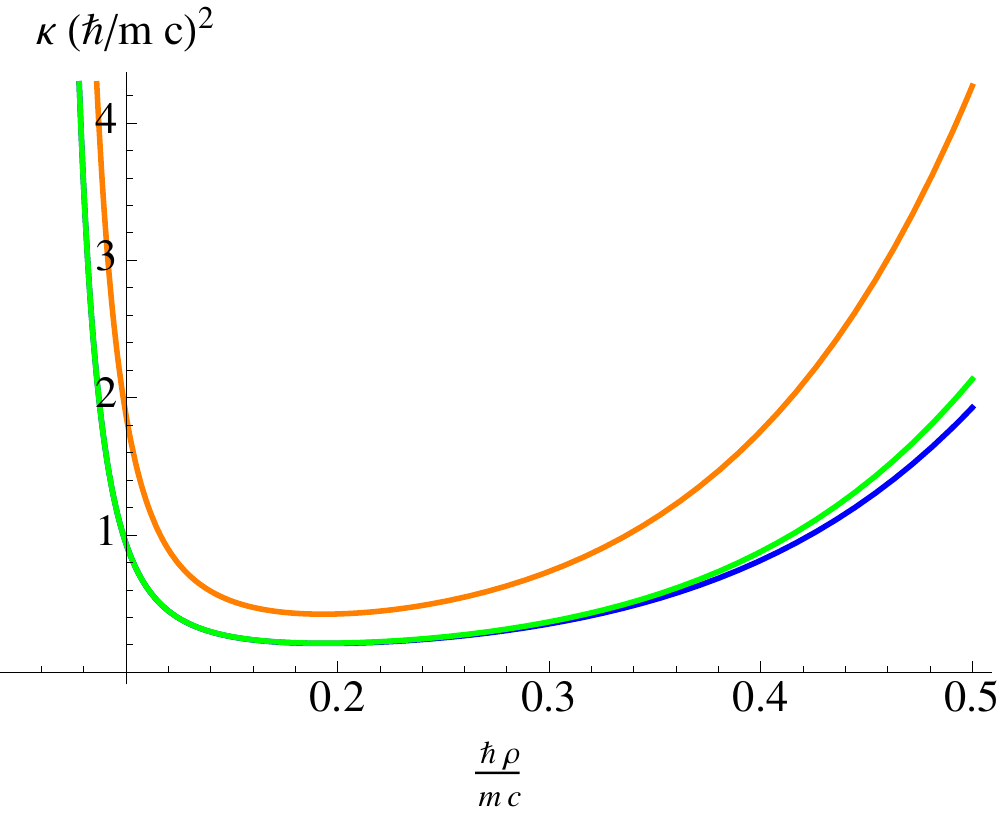}
\caption{\label{fig:kinkcnststrategies} (Color online) Comparison of the constant scaling factor $\kappa$ for different spacetime estimation strategies, $ \omega=1.01$ and $\varepsilon=0.1$ (in natural units with mass scale set by $m=1$). $\kappa$ is defined by the inequality $(\Delta \rho)^2\geq\frac{\kappa}{n N}$ where $n$ is the mean particle number of the initial state, and $N$ is the number of repetitions of the experiment.  (top orange) Coherent state with heterodyne measurements, (middle green) coherent state with homodyne measurements and (bottom blue) Fock state with particle counting measurements. For all values of the mass the Fock state strategy gives the lowest scaling factor and hence the best estimation of the kink curvature parameter, $\rho$. }
\end{figure}

\begin{figure}
\centering
\includegraphics[scale=.82]{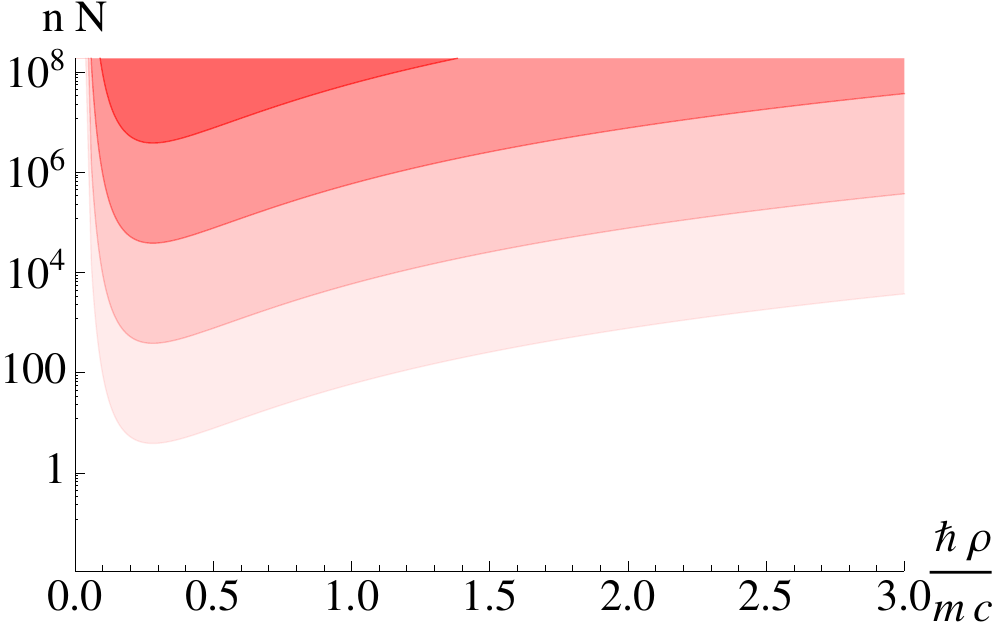}
\llap{\raisebox{.5cm}{
\includegraphics[scale=.45]{Blackhole_sensitivity_plot_legend}
}\hspace{.8cm}}
\caption{\label{fig:kinksens} (Color online) Relative sensitivity contour lines, $\epsilon=\left(\rho\Big|\frac{d\phi}{d\rho}\Big|\sqrt{4 n N}\right)^{-1}
$, as a function of the kink curvature parameter for $\omega\sim 1535$THz. $n$ is the particle number of the initial Fock state, and $N$ is the number of repetitions of the experiment. }
\end{figure}

We consider a single observer on the left side of the kink, who sends in probes to estimate the curvature parameter, $\rho$, from the reflected waves, see Fig.~\ref{fig:potkink}. Since the transmitted part of the signal propagates away from the observer it is effectively lost. Assuming that there is no radiation ingoing from the left hand side, and that the experimenter only makes measurements of the reflected wave in the right hand side, the problem becomes identical to that found for black holes \footnote{One stark difference between these two spacetimes is that in the black hole case there is a horizon. However, in the tortoise coordinate the horizon is moved all the way to negative infinity. So apart from the Hawking radiation, which we can safely neglect in the large frequency limit, the existence of the horizon is inconsequential to our analysis.} except that now the curvature parameter $\rho$ is being estimated instead of the mass. We can immediately infer that the channel is also a lossy bosonic channel given by (\ref{lossymap}) with $\cos{\phi}=\Big|\frac{\beta}{\alpha}\Big|=\sinh{(\frac{\pi k_-}{\rho})}/\sinh{(\frac{\pi k_+}{\rho})}$. Furthermore, Fock state preparation and particle counting will again give the optimal strategy. Results for the comparison of scaling constants and the relative sensitivity of the optimal Fock state measurement strategy are shown in Fig.~\ref{fig:kinkcnststrategies} and Fig.~\ref{fig:kinksens} respectively.

\section{Analogue black hole potentials in a waveguide}\label{sec:waveguide}
To get a feel for the magnitudes involved, the peak sensitivity in the black hole mass measurement occurs when the width of the potential is of the order of the wavelength of the probe field.  For optical wavelengths this corresponds to a black hole of about $10^{-10}M_{\odot}$. Such black holes are too small to be produced by astrophysical processes and too large to be produced in hypothesized extensions of the standard model \cite{miniblackholes}. While this method could be used to measure any sized black hole by appropriately choosing the probe frequency, in order to demonstrate our results we focus on settings suitable for quantum optics experiments. This puts the characteristic size of the analogue black holes that we consider at the order of microns.  

One easily verifies that our results do not depend critically on the curved spacetime background. Rather, the form of equations (\ref{masterequation}) and (\ref{kinkschrodinger}) and certain assumptions about the actions of the measurer and the initial state of the field lead immediately to the identification of the lossy channel (\ref{lossymap}). Recently there has been a lot of interest in constructing analogue experiments that reproduce the behavior of quantum fields in curved space-times  \cite{analogue}. This has been for the purpose of observing Hawking radiation. In our case, we are interested in the estimation of the black hole mass in the regime in which the Hawking radiation is negligible. Therefore, an analogue black hole potential for our purposes is simply a physical system having field equations of the Schr\"{o}dinger form (\ref{masterequation}) or (\ref{kinkschrodinger}). Any physical systems satisfying this equation will reproduce the relevant phenomena that we have investigated. From the many candidate systems available, those which provide excellent precision and control over the shape of the potential give the most accurate representations the actual space-times. 

It should be emphasized that in constructing the analogue system in this way there is strictly speaking no black hole. The tortoise coordinate has moved the horizon all the way to spatial infinity. Therefore, there is no region of the physical space from which light can not escape. Nevertheless, our objective is to model the potential in the external region of the Schwarzschild space outside the horizon since this allows us to investigate (in the regime of negligible Hawking radiation) the black hole scattering problem in analogue systems, our approach should therefore be distinguished from that of \cite{Schutzhold2005}.

\begin{figure}
\centering
\includegraphics[scale=.3]{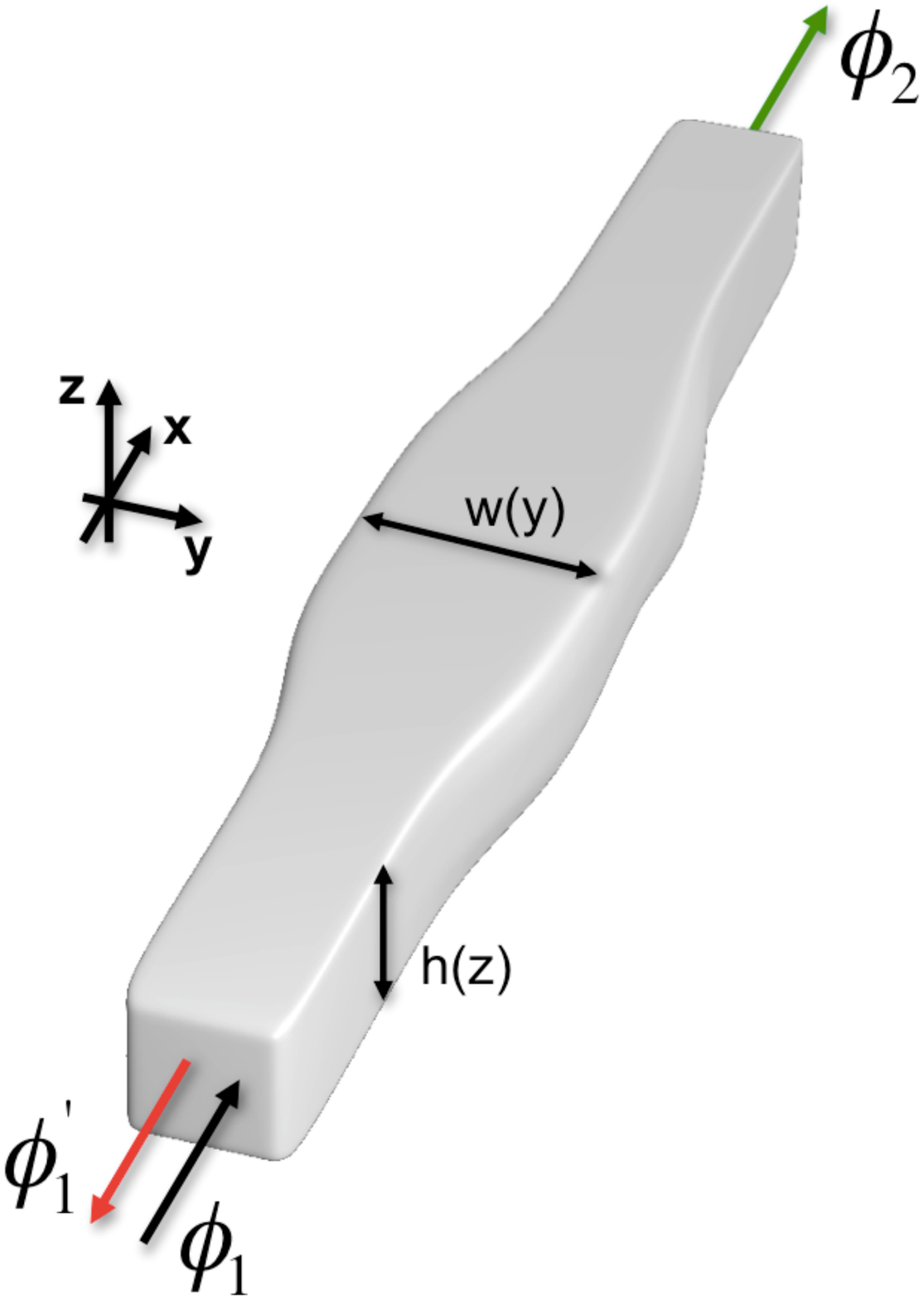}
\caption{\label{fig:waveguide} (Color online) Schematic diagram of the waveguide.  The electromagnetic wave in the $x$-direction. By modifying the geometry $w(y)$ and $ h(z)$ the effective index of refraction of the electromagnetic radiation can be precisely controlled and mapped onto the equations satisfied by a field in a static spacetime (see main text section \ref{sec:waveguide}).  }
\end{figure}

Waveguides with position dependent geometries provide promising realizations of such analogues \cite{zentgraf2011,Falco2011}, see also \cite{Dentcho, Sheng}.  Consider Maxwell's equations for an electric field with sinusoidal time dependence $e^{-i\omega t}$ that is polarized in the $z$-direction in a non-dissipative medium with permittivity $\varepsilon$ and permeability $\mu$:
\bea\label{Maxwell}
\left(\nabla^2+\mu\varepsilon\omega^2\right)E_z(x,y,z)=0.
\eea
We suppose that the mode propagates in the $x$-direction and that the transversal geometry of the waveguide varies along this direction, see Fig.~\ref{fig:waveguide}. The solution is assumed to remain in the fundamental mode of the transversal Laplacian $\nabla_{\perp}^2E_z=-\beta(x) E_z$ where the eigenvalue $\beta(x)$ depends on the geometry the waveguide. We expect this assumption to be good when the transversal geometry does not vary significantly over distances shorter than the wavelength $\beta^{-1}|\frac{d\beta}{dx}|\ll \lambda^{-1}$. Equation (\ref{Maxwell}) can then be written:
\bea\label{hemholtz}
\frac{d^2E_z(x,y,z)}{dx^2}+\left(\mu\varepsilon \omega^2-\beta(x)\right)E_z(x,y,z)=0.
\eea   
By identifying $E_z$ with $R$ (or $Z$) and arranging the geometry of the waveguide such that it matches the desired potential, (\ref{masterequation}) or (\ref{kinkschrodinger}), we obtain the required analogue equation. Note that the tortoise coordinate, $x$, becomes the position variable, $x$, in the analogue equation. 

A solution for the corresponding geometry can be found by determining the effective index of refraction along the $x$-direction. Using a WKB approximation, $E_z(x)=e^{S(x)}$, $\ddot{S}\ll\dot{S}^2$, one finds $\dot{S}=i\sqrt{\mu\varepsilon\omega^2-\beta(x)}$. Therefore the phase near the point $x_0$ is $\phi=\sqrt{\mu\varepsilon\omega^2-\beta(x_0)}x-\omega t$, giving a phase velocity:
\bea
v_p(x_0)=\frac{dx}{dt}\Big|_{x=x_0}&=&\frac{\omega}{\sqrt{\mu \varepsilon \omega^2 -\beta(x_0)}},
\eea
or an effective index of refraction:
\bea
n_{\text{eff}}(x)\equiv \frac{c}{v_p(x)}&=&\sqrt{\varepsilon_r}\sqrt{1-\frac{c^2\beta(x)}{\varepsilon_r\omega^2}},
\eea
where $\varepsilon_r\equiv\varepsilon/\varepsilon_0$ is the relative permittivity and $\varepsilon_0$ is the permittivity of free space. 

We now provide a mapping of the black hole potential onto a waveguide. The function $\beta(x)$ corresponds to the black hole potential. For the black hole potential (\ref{masterequation}), we have (putting back units of $c$):
\bea
\beta(x)=\left(1-\frac{2MG}{c^2r(x)}\right)\left(\frac{2M G}{ c^2r(x)^3}\right),
\eea
where we consider $s$-wave scattering in the actual black hole spacetime, i.e., $l=0$, and 
\bea
r(x)&=&\frac {2 M G}{c^2}  \left\{1+W\left(\exp\big({\frac{x c^2} {2 M G}-1}\big)\right)\right\},
\eea
where  $W$ is the Lambert W function.
 
The maximum of the gravitational potential occurs at
\bea
V_{\text{max}}=\frac{27c^4}{1024 M^2G^2},
\eea
which is mapped to the minimum index of refraction in the analogy: 
\bea
n_{\text{min}} \sim\sqrt{\varepsilon_r}\sqrt{1-\frac{27c^6}{1024 M^2G^2 \omega^2\varepsilon_r}}.
\eea
The maximum effective index of refraction occurs for the black hole at spatial infinity where the potential is zero, and is therefore given by the square root of the relative permittivity of the material filling the waveguide: $n_{\text{max}}=\sqrt{\varepsilon_r}$. At the optimal frequency (\ref{optimalomega}) the effective refractive index can be parameterized by the optimal wavelength ($\lambda_\text{opt}\equiv c/\omega_\text{opt}$) and is given by:
\bea
n_{\text{eff}}(x)&=&\sqrt{\varepsilon_r-\frac{0.3\lambda_\text{opt}^3}{ r(x)^3}\left(1-\frac{0.3\lambda_\text{opt}}{ r(x)}\right)},
\eea
where $r(x)=0.3\lambda_\text{opt}  \left\{1+W\left(\exp\big({\frac{x} {0.3\lambda_\text{opt}}-1}\big)\right)\right\}$.

\section{Conclusion}
We have investigated the measurability of gravitational parameters in static spacetimes using scattering experiments. While we focused on two examples, namely the Schwarzschild black hole and the spatially kinked metric, our approach could also be applied to a variety of other static spacetimes. 

Our setup consisted of considering a single observer, measuring the spacetime by sending probes through the gravitational potential and measuring the information contained in the reflected wave. We showed that because generically there is also transmission the channel is necessarily lossy. We identified the scaling limit to the measurement of the quantity parameterizing the potential and found that it obeys a shot-noise scaling relation. The best strategy for this type of scattering experiment was found to be Fock state probes and particle counting measurements. It should be noted that the results we have found set a limit only for the experimental arrangement we have considered. It is plausible that alternative experimental configurations can provide better performance c.f \cite{Downes}.    

While it would not be possible to perform optical experiments with real black holes, we showed that waveguides with position dependent geometries provide a very good experimental laboratory in which analogue black hole potential experiments could be conducted. It is worth mentioning that the tools of quantum enhanced metrology that we have applied here to the study of gravitational systems, could even be used for the more practical application of improving the characterization of fabricated waveguides. To our knowledge, quantum sensitivity has not been employed thus far for this purpose. The identification of Fock states as the ultimate resource for such characterization is particularly promising in view of recent progress in Fock state production \cite{fock} and improvements in photon counting techniques \cite{photoncounting}.

While we have focussed on the implementation of our spacetime measurement strategy in waveguide systems other analogue systems could also be used to investigate the quantum imposed limitations on the measurability of spacetime.  For example, there is a well-known correspondence between the propagation of electromagnetic fields in curved spacetimes and in dielectric media \cite{Plebanski}. Recent theoretical \cite{Dentcho} and very exciting experimental work \cite{Sheng} have produced analogue black holes in which the scattering problem that we have described appears to be feasible. Recent results \cite{Tuckerrefs} on the quantization of Maxwell's equations in epsilon-near-zero meta-materials with anisotropic and inhomogeneous permittivity also provide an interesting avenue in which the metrological tools we have discussed could be investigated. 

\appendix
\section{}\label{practicalFisher}

Here we provide details of the sub-optimal Gaussian strategies appearing in FIG.~\ref{fig:cnststrategies} and FIG.~\ref{fig:kinkcnststrategies}.  Since everything is Gaussian the calculation is simplified by working in the covariance matrix formalism. The optimal Gaussian strategy for the lossy bosonic channel has been found in \cite{Monras2007}. Because the ultimate optimal strategy is non-Gaussian we are more interested here in comparing it to simple and practical Gaussian strategies. Therefore we calculate the Fisher information for coherent and squeezed input states that are measured in homodyne and heterodyne. 
 
We write the action of a general beam splitter on the annihilation operators as:
\bea
\hat{a}_1'&=&R \hat{a}_1 +T \hat{a}_2; \\
\hat{a}_2'&=&-T^*\hat{a}_1+R \hat{a}_2,
\eea
where $R$ and $T$ are the reflection and transmission amplitudes respectively, $|T|=\sqrt{1-R^2}$, and $R$ is assumed to be real.

Recall we consider the situation where the transmitted mode is lost and the incoming mode from the far side of the potential is in the vacuum state which leads to a non-unitary single mode Gaussian map. Any single mode non-unitary Gaussian transformation can be expressed in terms of its action on the state's first moments $d_i=\langle \hat{x}_i\rangle$ and covariance matrix $\sigma_{ij}=\tfrac{1}{2}\langle \hat{x}_i\hat{x}_j+\hat{x}_j\hat{x}_i\rangle+\langle \hat{x}_i\rangle\langle\hat{x}_j\rangle$ with $i\in \{x,p\}$, according to \cite{Serafini}:
\begin{align} \label{eq1}
d^{\hspace{1 pt}\text{out}}&=X\hspace{1 pt}d^{\hspace{1 pt}\text{in}},\\
\sigma^{\text{out}}&=X\hspace{1 pt}\sigma^{\text{in}}\hspace{1 pt}X^{T}+Y, \label{eq2}
\end{align}
where we define the quadratures $\hat{x}_x\equiv\hat{a}_1'+\hat{a}'_1{}^{ \dagger}$ and $\hat{x}_p\equiv\frac{1}{i}\left(\hat{a}'_1-\hat{a}'{}^{\dagger}_1\right)$. We find $X=R\hspace{1 pt}I_2$ and $Y=(1-R^2)I_2$, where $I_2$ is the $2\times2$ identity matrix.

The first measurement we consider is Homodyne detection with a coherent state input. The first moments and covariance matrix for this input are $d^{\hspace{1 pt}\text{in}}=2\{\text{Re}(\alpha),\text{Im}(\alpha)\}$ and $\sigma^{\hspace{1 pt}\text{in}}=I_2$, where $\alpha$ is the displacement. Using (\ref{eq1}, \ref{eq2}) the output state has $d^{\hspace{1 pt}\text{out}}=2R\{\text{Re}(\alpha),\text{Im}(\alpha)\}$ and $\sigma^{\text{out}}=I_2$, which again is a coherent state, $\left|R\alpha\right>$. We wish to calculate the optimal measurement w.r.t. the Fisher information, which for a homodyne measurement means we must maximize the Fisher information over all possible measurement quadratures $\hat{x}({\theta})\equiv\cos\theta \hat{x}_x+\sin\theta \hat{x}_p$, $\hat{p}({\theta})\equiv\cos\theta \hat{x}_p-\sin\theta \hat{x}_x$. We begin by finding the probability density for the results of homodyne measurements, $p\big(x({\theta})|\rho^{\text{out}}(R)\big)=\int_{\mathbb{R}}W\big(x({\theta}),p({\theta})\big)dp({\theta}),$ where $W$ is the Wigner function of the output state: 
\begin{align}
W\big(x({\theta}),p({\theta})\big)=\frac{1}{2\pi}\exp{\left(-\frac{\tilde{x}(\theta)^2+\tilde{p}(\theta)^2}{2}\right)},
\end{align}
where we have defined $\tilde{x}(\theta)\equiv x(\theta)-\left<\hat{x}(\theta)\right>$ and $\tilde{p}(\theta)\equiv p(\theta)-\left<\hat{p}(\theta)\right>$. We find:
\beq
p\big(x({\theta})|\rho^{\text{out}}(R)\big)=\frac{1}{\sqrt{2\pi}}\exp\left(-\frac{\tilde{x}({\theta})^2}{2}\right).
\eeq
We find that the Fisher information (\ref{FI}) is given by $F=4\left(\text{Re}(\alpha)\cos\theta+\text{Im}(\alpha)\sin\theta\right)^2$. The maximum for this function is at $\cos\theta=\text{Re}(\alpha)/\left|\alpha\right|$, $\sin\theta=\text{Im}(\alpha)/\left|\alpha\right|$, for which $F=4\left|\alpha\right|^2=4n$, where $n$ is the mean particle number in the coherent state. We note that although there is no dependence on $R$, one usually wants to estimate some other parameter on which the reflection amplitude depends. For example, the Fisher information in terms of the black hole mass, $F(M)$, or the spacetime curvature parameter, $F(\rho)$, are found using the reparameterization property and are given by: $F(R(M))=4n (\frac{dR}{dM})^2$ and $F(R(\rho))=4n (\frac{dR}{d\rho})^2$ respectively.

Moving now to heterodyne measurements of initial coherent states. The measurement is given by the set of projectors $\frac{1}{\pi}\{|\beta\rangle\langle\beta|\}$ where $|\beta\rangle$ are the set of single mode coherent states. We consider an initially coherent probe state, which without loss of generality we consider to be displaced in the $x$ direction, i.e., $|\alpha\rangle$ with $\alpha$ real.  Since coherent states will remain coherent under the lossy bosonic channel the reflected output will also be coherent and is found to be $|R\alpha\rangle$. The probability distribution is then obtained from the overlap of two coherent states. This is easily calculated using the formula for the overlap of two arbitrary Gaussian states. We find:
\bea
p(\beta|R)=\frac{1}{\pi}\exp \big(- (R\alpha-\text{Re}{\beta})^2-(\text{Im}{\beta})^2\big),
\eea
from which we obtain the Fisher Information $F=2\alpha^2=2n$, which is exactly half that found for homodyne measurements. 

\acknowledgments
We thank G. Adesso, J. Louko and A. Dragan for useful discussions. J.D and I. F. acknowledge support from EPSRC (CAF Grant No. EP/G00496X/2 to I. F.). A.D.F and D.F. acknowledge support from EPSRC ( EP/J004200/1). D.F. acknowledges financial support from the European Research Council under the European Union's Seventh Framework Programme (FP/2007-2013)/ERC GA 306559 and EPSRC (UK, Grant EP/J00443X/1).

\end{document}